\documentclass[english]{iopart}
\usepackage{graphicx}
\usepackage{iopams,setstack}
\usepackage{babel}
\usepackage{verbatim} 

\newcommand{\bp}{{\bf p}}
\newcommand{\bq}{{\bf q}}

\newcommand{\beq}{\begin{equation}}
\newcommand{\eeq}{\end{equation}}

\newenvironment{lyxlist}[1]
{\begin{list}{}
{\settowidth{\labelwidth}{#1}
 \setlength{\leftmargin}{\labelwidth}
 \addtolength{\leftmargin}{\labelsep}
 }}
{\end{list}}
 
\begin{document}

\title{Metastability in spin polarised Fermi gases and quasiparticle 
decays}

\author{K Sadeghzadeh$^1$, G M Bruun$^2$, C Lobo $^{1,3}$, P Massignan$^{4,5}$ and A Recati$^6$}

\address{$^1$ Department of Physics, Cavendish Laboratory, JJ Thomson Avenue, Cambridge, CB3 0HE, UK}
\address{$^2$ Department of Physics and Astronomy, Aarhus University, Ny Munkegade 120, DK-8000 Aarhus C, Denmark}
\address{$^3$ School of Mathematics, University of Southampton, Highfield, Southampton, SO17 1BJ, UK}
\address{$^4$ F\'isica Teorica: Informaci\'o i Processos Qu\`antics, Universitat Aut\`onoma de Barcelona, 08193 Bellaterra, Spain}
\address{$^5$ ICFO-Institut de Ci\`encies Fot\`oniques - Mediterranean Technology Park, 08860 Castelldefels (Barcelona), Spain}
\address{$^6$ INO-CNR BEC Center and Dipartimento di Fisica, Universit` di Trento,via Sommarive 14, I-38123 Povo, Italy}
\ead{ks539@cam.ac.uk}

\begin{abstract}We investigate the metastability 
associated with the first order transition from normal to superfluid phases 
in the phase diagram of two-component  polarised Fermi gases.
We begin by detailing the dominant decay processes of single quasiparticles.
Having determined the momentum thresholds of each process and calculated their rates, we apply this understanding to a Fermi sea of polarons by linking its metastability to the stability of individual polarons, and predicting a region of metastability for the normal partially polarised phase.
In the limit of a single impurity, this region extends from the interaction strength at which a polarised phase of molecules becomes the groundstate, 
to the one at which the single quasiparticle groundstate changes character from polaronic to molecular.
Our argument in terms of a Fermi sea of polarons naturally suggests their use as an experimental probe.
 We propose experiments to observe the threshold of the predicted region of metastability, the interaction strength at which the quasiparticle groundstate changes character, and the decay rate of polarons.
\end{abstract}

\submitto{\NJP Focus issue on Strongly Correlated
Quantum Fluids: From Ultracold Quantum Gases to QCD Plasmas}
\maketitle 
 
\section{Introduction}
 
The experimental realisation of spin polarised ultracold Fermi gases has 
initiated a variety of new physics 
\cite{PhysRevLett.102.230402,PhysRevLett.103.170402,PhysRevLett.97.030401,PhysRevLett.97.190407,N.Navon05072010,Partridge01272006,Zwierlein01272006,nature06473,nature08814}. 
Of particular interest is the understanding of strongly interacting two-component Fermi 
gases at zero temperature. Two theoretical approaches have been used to shine light on this intriguing problem: the study of a single atom immersed in an ideal Fermi gas of atoms in a different spin state and, a Monte Carlo calculation at finite polarization. The Monte Carlo approach revealed the theoretical 
phase diagram of a homogeneous Fermi gas, as a function of polarisation $P=(N_\uparrow-N_\downarrow)/(N_\uparrow+N_\downarrow)$
and interaction strength \cite{PhysRevLett.100.030401}, mapping out a phase 
separation of superfluid and normal phases. The single impurity approach 
examines the quasiparticles used as the building blocks to describe these 
phases. In the strongly imbalanced limit, a single $\downarrow$ fermion 
immersed in a Fermi sea of $\uparrow$ fermions, the spin impurity atom becomes 
either a fermionic (polaron) or bosonic (molecule) quasiparticle. 
Complementary wave functions for each quasiparticle 
(\cite{PhysRevA.74.063628,PhysRevLett.98.180402} and \cite{PhysRevA.80.053605,PhysRevA.80.033607,Europhys}, respectively) provide groundstate energies and effective masses. 
Previous studies showed that the critical interaction strength at
which the
groundstate of a single impurity at zero momentum switches from the
fermionic to
bosonic branch \cite{PhysRevB.77.020408} occurs at ${1}/{(k_{F\uparrow}a)_{c}}\backsim0.88$, with $k_{F\uparrow}=(6\pi^2n_\uparrow)^{1/3}$ the Fermi momentum of a non-interacting Fermi sea of $\uparrow$ atoms with density $n_\uparrow$, and $a$ the scattering length parametrizing the
interaction strength between $\uparrow$ and $\downarrow$ atoms. This value is higher than
the interaction strength ${1}/{k_{F\uparrow}a}\backsim0.73$ at which a superfluid
phase emerges in the limit of full polarisation \cite{PhysRevLett.100.030401}.
An important conclusion made from this is that no measurement made in the groundstate will allow us to see the point at ${1}/{(k_{F\uparrow}a)_{c}}$. As we will show later the use of metastable and out of equilibrium processes overcomes this problem.
 
Firstly, we determine the momentum thresholds for the decay of single
impurities with
finite momentum. We then consider the metastability associated with the
normal to superfluid first order phase transition in the thermodynamic
phase diagram. Indeed by applying the understanding of single
quasiparticle decay to a Fermi sea of polarons, we predict that there
exists a region where such a phase is metastable. We propose the use of a Fermi sea
of polarons as an 
experimental probe to determine the threshold of the 
region. As this threshold goes to zero at 
${1}/{(k_{F\uparrow}a)_{c}}$, we can observe this point for the first 
time. Beyond this threshold the decay of polarons into molecules may lead to a mixture of molecules and polarons, which suggests a possible way of measuring the 
molecule-polaron scattering length. In this way, we hope to open up a new 
regime of metastable physics in Fermi gases for experimental exploration. 
Finally, we calculate the decay rates for each process, within the key 
regions of interaction strength and momenta, to determine the fate of the 
quasiparticles. The presence of a Fermi sea of
polarons would again be instrumental to measure the various decay
rates.

\section{Background}
 
\begin{figure}
\begin{center}
\includegraphics[width=\columnwidth]{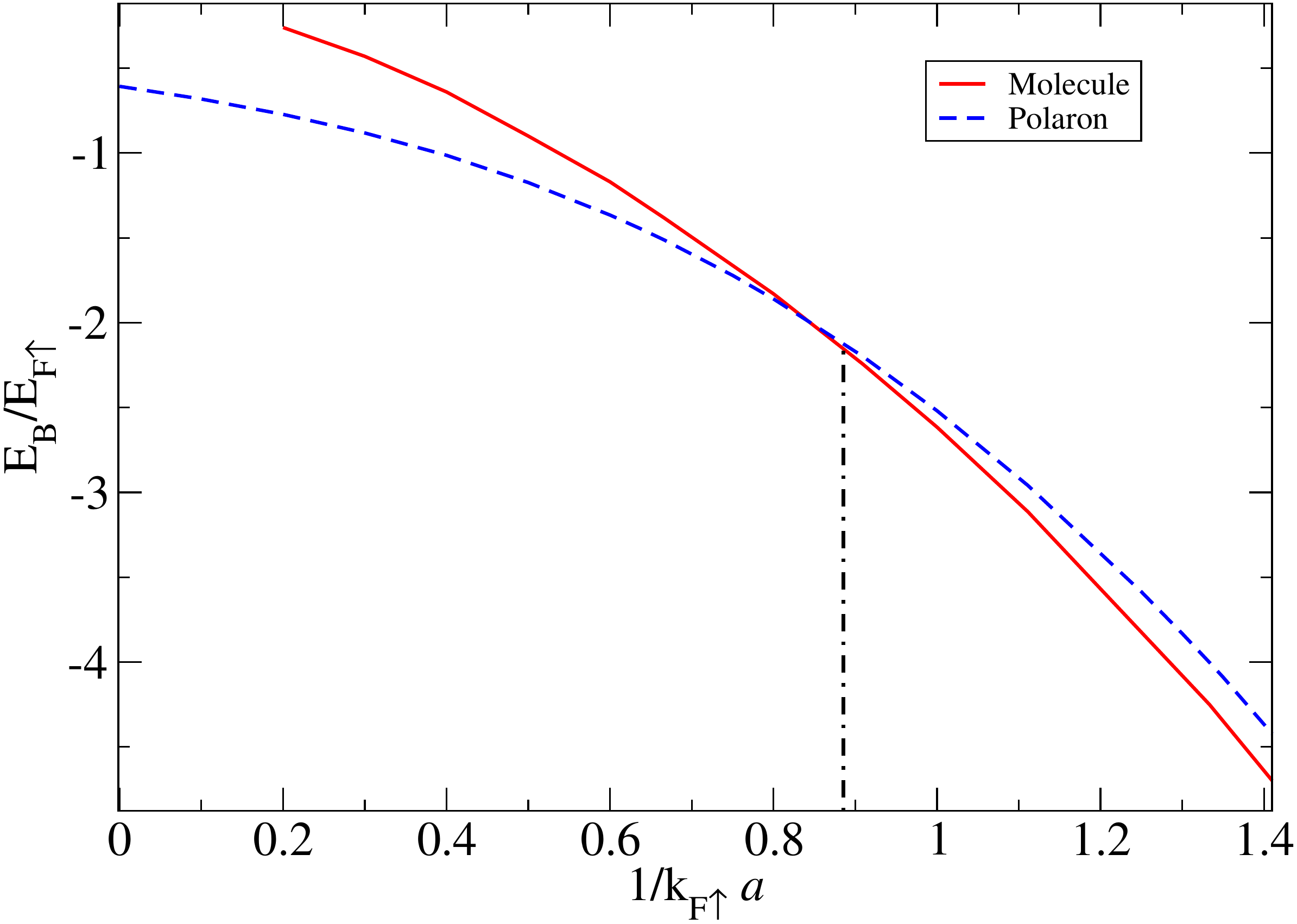}
\caption{{\footnotesize Molecule and polaron 
zero momentum energies as a function of interaction strength on the 
{}``BEC'' side. Red line: Molecule. Blue dashed line: Polaron. The dot-dashed line 
marks the critical interaction strength 
${1}/{(k_{F\uparrow}a)_c}\backsim0.88$. To the left of it the polaron energy 
is lower, and to the right it is the molecule that becomes the 
groundstate.}}
\label{fig:BEC-BCS-crossover}
\end{center}
\end{figure}

Our starting point is an understanding of the single quasiparticle 
groundstate as a function of ${1}/{k_{F\uparrow}a}$ going from 
unitarity (${1}/{k_{F\uparrow}a}\rightarrow0$) to the {}``BEC'' limit 
(${1}/{k_{F\uparrow}a}\rightarrow+\infty$) (see 
figure \ref{fig:BEC-BCS-crossover}).  At ${1}/{(k_{F\uparrow}a)_{c}}\sim0.88$, 
the critical point$ $, the zero momentum energies of the polaron and 
molecule cross. For smaller values of 
${1}/{k_{F\uparrow}a}$ the polaron is the groundstate of the system, 
and for larger values the molecule is the groundstate. In figure 
\ref{fig:BEC-BCS-crossover} and throughout this paper we use the polaron 
energy calculated using the wave function proposed by Chevy 
\cite{PhysRevA.74.063628}. The energy in the BCS limit tends to 
$E_{Pol}={4\pi a n_{\uparrow}}/{m}$, the mean-field interaction energy, and 
in the BEC limit tends to 
$E_{Pol}=-\frac{1}{ma^{2}}-\frac{\epsilon_{F\uparrow}}{2}$. Here and in the following, we take $\hbar=1$.
 The effective mass of a polaron ($m_{Pol}^{*}$) obtained 
from \cite{Europhys} becomes just the bare mass in the BCS limit and 
diverges at ${1}/{k_{F\uparrow}a}\sim1.17$. For the molecule, energies 
and effective masses are obtained from 
\cite{PhysRevA.80.053605,PhysRevA.80.033607,Europhys}. In the BEC limit the 
energy tends to $E_{Mol}=-\frac{1}{ma^{2}}-\epsilon_{F\uparrow}$. The 
effective mass equals the bare molecule mass $m_{Mol}^{*}=2m$ in that limit, while approaching unitarity it grows,
and it diverges at ${1}/{k_{F\uparrow}a}\sim0.55$. These masses and 
energies for both polarons and molecules have been found to be very 
accurate by comparison with Quantum Monte Carlo calculations \cite{PhysRevLett.100.030401,
PhysRevB.77.020408}.

\section{Decay processes and thresholds}

We begin by considering the decays that the polarons and molecules can undergo when they are not in the groundstate (e.g. at finite momentum). The decay process of a quasiparticle can only occur when it satisfies energy and momentum conservation. This is generally only possible when the initial quasiparticle momentum $p$ is above a threshold momentum $P_{Th}$.

\subsection{Polaron decay}\label{PolaronDecay}
 For the polaron at zero temperature, the decay processes we consider are:
\begin{lyxlist}{00.00.0000}
\item [{A.}] $Polaron\rightarrow Molecule+hole$
 \item [{B.}] $Polaron\rightarrow Molecule+2\; holes+particle$
\end{lyxlist}
A finite momentum polaron also has a finite relaxation time as described
in \cite{PhysRevLett.100.240406}, which schematially reads
\begin{lyxlist}{00.00.0000}
 \item [{C.}] $Polaron\left(p\right)\rightarrow Polaron\left(p'<p\right)+hole+particle$ 
\end{lyxlist}

The three-body decay of a polaron into a molecule, two holes and a particle has been considered in \cite{PhysRevLett.105.020403} in the special case $\bp=0$. Generalizing this calculation to non-zero momentum, we find that process B can occur at any momentum when $\Delta E=E_{Pol}-E_{Mol}>0$, as expected, and only for 
 \beq
 p>P_{Th}^{B}=\sqrt{-2m_{Pol}^{*} \Delta E }
 \label{PthB}
 \eeq
  when $\Delta E<0$. Here $E_{Pol}$  and $E_{Mol}$ are the energies of a zero-momentum polaron and molecule at the given value of interaction strength.

For process A, the two-body decay of a polaron with momentum $\bp$ into a molecule and a hole, conservation of energy and momentum require the following equality to be verified:
\beq
E_{Pol}+\frac{p^2}{2m^*}+\xi_q=E_{Mol}+\frac{(\bp+\bq)^2}{2M^*},
\eeq
where $\xi_{q}=\frac{q^{2}}{2m}-\epsilon_{F\uparrow}$ is the kinetic energy of a 
majority particle measured with respect to the Fermi surface. The minimum momentum $P_{Th}^{A}$ at which process A is allowed is found by setting the hole at the Fermi surface, and by taking its momentum to be anti-parallel to the one of the polaron:
\begin{equation} 
\frac{P_{Th}^{A}{}^{2}}{2m_{Pol}^{*}}+E_{Pol}=\frac{\left(P_{Th}^{A}-p_{F}\right)^{2}}{2m_{Mol}^{*}}+E_{Mol}.
\label{PthA}
\end{equation}

The processes including more particle-hole pairs (e.g. $Molecule+3\; holes+2\; 
particles$) have the same energy threshold as the state resulting from 
process B, but lower rates so we do not consider them here.

In conclusion, polarons with non-zero momentum are stable towards A/B decay for momenta smaller than $P_{Th}^{A/B}$.

Figure \ref{fig:momentum threshold} shows our results for the polaron decay thresholds, as given by (\ref{PthB}) and (\ref{PthA}).
In the region below ${1}/{(k_{F\uparrow}a)_{c}}$, where the zero momentum 
polaron is the groundstate of the system, a polaron at finite momentum can nevertheless 
be unstable to decay processes A, B, and C. The solid blue line 
gives the momentum threshold $P_{Th}^{B}$ for the polaron to decay via process B. On this line, 
 decay results in a zero momentum molecule for ${1}/{(k_{F\uparrow}a}<{1}/{(k_{F\uparrow}a)_{c}}$. Finite momentum 
molecules also result from process B above the solid blue line. The blue dashed line gives the momentum threshold $P_{Th}^{A}$
for process A, which 
generally creates a molecule at finite momentum even on the threshold. 
Decay processes including more particle-hole pairs all set in at momenta above the solid blue line, i.e.\ for $p>P_{Th}^{B}$. For ${1}/{(k_{F\uparrow}a}>{1}/{(k_{F\uparrow}a)_{c}}$, a polaron at 
zero momentum is unstable to a process B decay into a molecule at finite 
momentum whereas process A continues to affect only higher momentum polarons. 
Polarons with any finite momentum are unstable to momentum relaxation 
(process C) on both sides of ${1}/{(k_{F\uparrow}a)_{c}}$.
The momentum thresholds for processes A and B become equal where the molecule's effective mass diverges 
(${1}/{k_{F\uparrow}a}\sim0.55$).

\subsection{Molecule decay}
 The stability of a molecule is calculated in a similar way. The decay 
channels we consider are: 
\begin{lyxlist}{00.00.0000} \item [{$\tilde{A}$.}] $Molecule\rightarrow 
Polaron+particle$ \item [{$\tilde{B}$.}] $Molecule\rightarrow Polaron+2\; 
particles+hole$ \item [{$\tilde{C}$.}] $Molecule\left(p\right)\rightarrow 
Molecule\left(p'<p\right)+hole+particle$ \end{lyxlist} 
For
$1/k_Fa<{1}/{(k_{F\uparrow}a)_{c}}$, a zero momentum molecule decays via process 
$\tilde{B}$ to a polaron with finite momentum, and via process $\tilde{A}$ 
at higher momentum. For 
$1/k_Fa>{1}/{k_{F\uparrow}a_{c}}$, process $\tilde{B}$ preceeds process 
$\tilde{A}$ with increasing momentum until where the polaron's effective 
mass diverges.
In analogy with the polaron decay considered
above, the momenum thresholds for the molecule are determined by
energy and momentum conservation. They are,
 \begin{equation} 
\frac{P_{Th}^{\tilde{A}}{}^{2}}{2m_{Mol}^{*}}+E_{Mol}=\frac{\left(P_{Th}^{\tilde{A}}-p_{F}\right)^{2}}{2m_{Pol}^{*}}+E_{Pol},
\label{PthtildeA}
\end{equation}
\begin{equation} 
P_{Th}^{\tilde{B}}=\sqrt{2m_{Mol}^{*} \Delta 
E}\quad 
\rm{for}\;\frac{1}{\it{k_{F\uparrow}a}}>\frac{1}{\it{(k_{F\uparrow}a)_{c}}},
\label{PthtildeB}
\end{equation}
 where $P_{Th}^{\tilde{B}}=0$ for 
$1/k_{F\uparrow}a\leqslant1/{(k_{F\uparrow}a)_{c}}$. Again, the threshold 
momentum for process $\tilde{A}$ is found when the majority particle is on the Fermi surface.

\begin{figure}
\begin{center}
\includegraphics[width=\columnwidth]{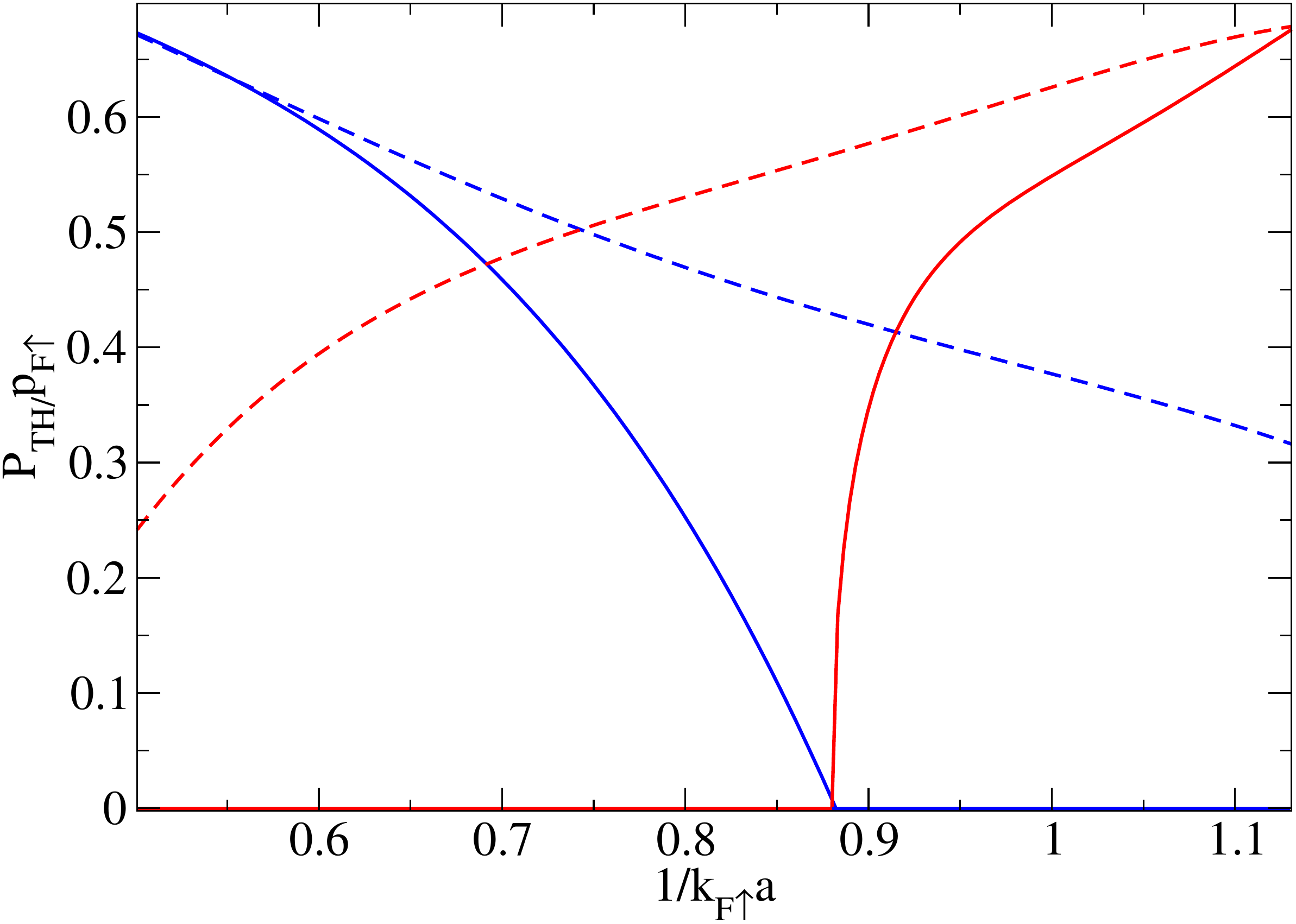}
\caption{{\footnotesize Momentum thresholds 
for various decay processes for both polarons and molecules about the 
critical point. Blue solid line: $Polaron\rightarrow Molecule+2\; 
holes+particle$; Blue dashed line: $Polaron\rightarrow Molecule+hole$; Red solid line: 
$Molecule\rightarrow Polaron+2\; particles+hole$; Red dashed line: 
$Molecule\rightarrow Polaron+particle$. The groundstate of the system changes from being a polaron to a molecule at 
${1}/{k_{F\uparrow}a}\sim0.88$, beyond which even a polaron at $p=0$ can 
decay via process B; otherwise, only polarons at $p>0$ are unstable in this 
way. Process A is relevant at higher polaron momentum and generally results 
in a finite momentum molecule even on the threshold. The process A and B 
thresholds meet when the molecule's (polaron's) effective mass diverges at 
${1}/{k_{F\uparrow}a}\sim0.55\left(1.17\right)$. Note that for polarons at large $p\left(\sim p_{F\uparrow}\right)$, 
momentum relaxation processes are very strong so that such a polaron is no 
longer a well-defined quasiparticle.}}
\label{fig:momentum threshold}
\end{center}
\end{figure}

Our results for the threshold momenta for polaron and molecule decay, as given by (\ref{PthB}), (\ref{PthA}), (\ref{PthtildeA}) and, (\ref{PthtildeB}) are shown in figure \ref{fig:momentum threshold}.

\section{Metastability of polaron gas}
 Thus far we have analysed the decay processes of single quasiparticles.
The single quasiparticle problem is a limiting case of the spin imbalanced
Fermi gas. The groundstate phase diagram of a spin imbalanced Fermi
gas, with polarisation versus interaction strength, has been calculated
in \cite{PhysRevLett.100.030401} using a Monte Carlo approach. In this calculation, 
the $\downarrow$ atom concentration is kept finite for a macroscopic system even in the $P\rightarrow1$ limit.
We now examine how the stability analysis above is connected to
this equilibrium phase diagram. 
 
\begin{figure}
\begin{center}
\includegraphics[width=\columnwidth]{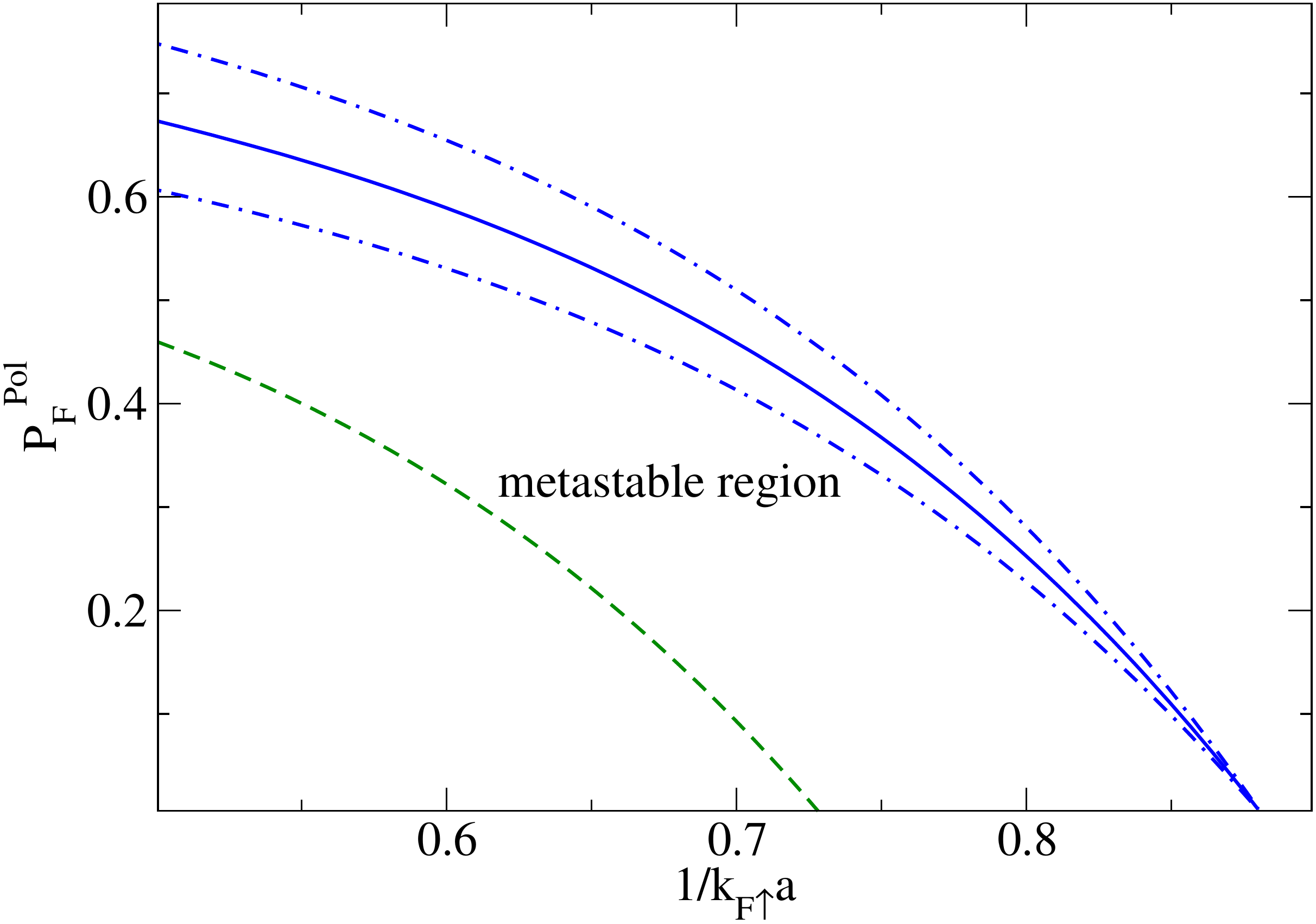}
\caption{{\footnotesize Stability of Fermi sea of 
polarons: The dashed green line shows the first order transition from normal 
to phases including a superfluid (Fig. 4 in \cite{PhysRevLett.100.030401}). 
The solid blue line is the Fermi momentum of a Fermi sea of polarons equal to 
the momentum threshold for process B ($p_{F}^{Pol}=P_{Th}^{B}$). The dot-dashed 
blue lines represent how the solid blue line might be shifted due to molecule-polaron 
interactions.}}
\label{fig:Phase-diagram}
\end{center}
\end{figure}

Our particular consideration is the normal to superfluid first order phase 
transition predicted by the Monte Carlo calculations \cite{PhysRevLett.100.030401}. 
This transition from a partially polarised normal phase to a state with  separated superfluid and normal phases 
 is represented by the dashed green line in figure \ref{fig:Phase-diagram}.
To make this line in figure \ref{fig:Phase-diagram}, we have assumed that the polarons form a weakly interacting 
Fermi sea, so that we can convert the critical density 
at which phase separation occurs, into a critical Fermi momentum  $p_{F}^{Pol}\left({1}/{k_{F\uparrow}a}\right)$ for the polarons. 
 In the phase separated state, the Monte Carlo calculation includes the interactions between the molecules in 
the condensate. In contrast, in the single $\downarrow$ atom calculation there is at most one molecule and therefore no condensate.

If we assume that the relevant processes in which the polarons can be converted into molecules, 
within experimentally realistic timescales, are only the 
ones considered in section \ref{PolaronDecay} and that these single impurity processes can be used to analyse the stability of the Fermi sea of polarons (neglecting for instance multiple polaron decay),
 this gives rise to a region of \emph{metastability} in the phase diagram. 
If a Fermi sea of polarons prepared in the groundstate
(below the green dashed line) is adiabatically taken above the dashed green line by increasing $1/k_Fa$,
we expect it to persist as a metastable state. Even though the density of the $\downarrow$ atoms is so high that the true groundstate 
is a phase separated state, the state is stable since 
there are no polarons with large enough momentum to decay to a molecule via process B. 
The decay of polarons to molecules sets in
only when $1/k_Fa$ is increased further so that the
Fermi momentum of the polarons is larger than the momentum threshold for process B 
($p_{F}^{Pol}=P_{Th}^{B}$); we ignore here molecule-polaron interactions (see below). This momentum is 
 given by the solid blue line in figure \ref{fig:Phase-diagram}. At this point, the highest momentum polarons will decay into molecules.

This motivates the following proposal to use a Fermi sea of polarons
as an experimental probe to explore the metastable region and observe
molecule formation through process B. Due to the Pauli exclusion principle, the momentum relaxation process (C) is suppressed 
allowing for an analysis of the other decay processes. Let us suppose for 
the moment that the molecules into which polarons can decay do not interact 
with the polarons. Then, to determine the shape of $P_{Th}^{B}$, we can 
prepare a polaron Fermi sea with a Fermi momentum smaller than the dashed 
green line in figure \ref{fig:Phase-diagram} at unitarity, for example. Then we 
increase ${1}/{k_{F\uparrow}a}$ adiabatically. The system will cross 
the first order phase transition line, but since it is metastable (as 
discussed above), it will remain a Fermi sea of polarons rather than form a 
condensate of molecules. As we continue beyond the phase transition, the 
Fermi momentum will become equal to the momentum threshold for process B 
($p_{F}^{Pol}=P_{Th}^{B}$). Increasing ${1}/{k_{F\uparrow}a}$ 
infinitesimally beyond this point will now lead to the decay of polarons at 
$p_{F}^{Pol}$.
The polarons on the Fermi surface decay into molecules at zero 
momentum. The appearance of these molecules can then be detected experimentally
as the tell-tale sign of the solid blue line $P_{Th}^{B}$.
The experiment can be repeated using an appropriate $p_{F}^{Pol}$ to find
$P_{Th}^{B}$ at different interaction strengths, making sure that only a small number of molecules are created each time
(so as to be able to ignore effects beyond the threshold, e.g. molecule-molecule interactions)
but a large enough number to be observable.
The size of the polaron Fermi sea needed will decrease as $P_{Th}^{B}$ decreases, so that the value $1/(k_{F\uparrow}a)_{c}$ is found in the limit of a single polaron.

If we now take into account molecule-polaron interactions the value of $p_{F}^{Pol}$ at which molecules are first formed will change. If we consider one 
molecule in the final state, the (unknown) molecule-polaron interaction 
changes the energy by $\Delta E=g_{PM}n_{Pol}$ (assuming a mean-field approximaton).
This will lead to a positive or negative shift 
of the threshold curve of production of molecules (see dot-dashed blue lines in figure 
\ref{fig:Phase-diagram}). This shift tends to zero at 
${1}/{(k_{F\uparrow}a)_{c}}$ where $n_{Pol}\rightarrow0$ and where the 
threshold curve meets $P_{Th}^{B}$. One could in principle use the 
difference between the experimentally observed curve and $P_{Th}^{B}$ 
(which is known theoretically) to determine this shift and so the 
molecule-polaron scattering length. Note that we have ignored the effects 
of the polaron-polaron interaction, which are known to be small.
 
We have therefore established that the region between the dashed green and
dot-dashed blue lines in figure \ref{fig:Phase-diagram} represents 
a metastable phase consisting of a Fermi sea of polarons.
Such a metastable phase is characteristic of a first order phase transition.
It is important to note that the state
may well be metastable beyond the dot-dashed blue line since an analysis of
its stability in that region would require us to take into account
the presence of a finite quantity of molecules. 
We also raise the intriguing possibility 
of a final state containing the remaining Fermi sea of polarons coexisting 
with a condensate of molecules, within a background of $\uparrow$ 
particles.

\section{Single quasiparticle decay rates and experimental observability}

\subsection{Decay rates calculation}

\begin{figure}
\begin{center}
\includegraphics[width=\columnwidth]{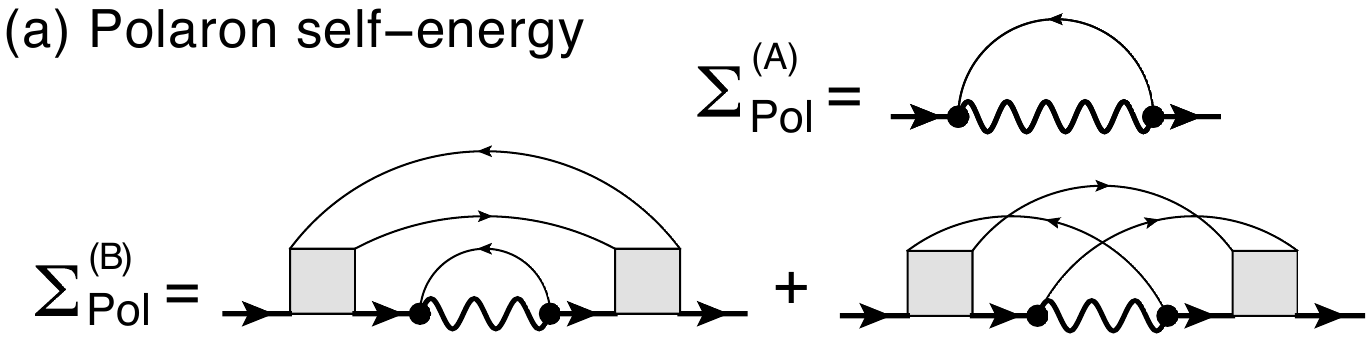}

\vspace{2cm}
\includegraphics[width=\columnwidth]{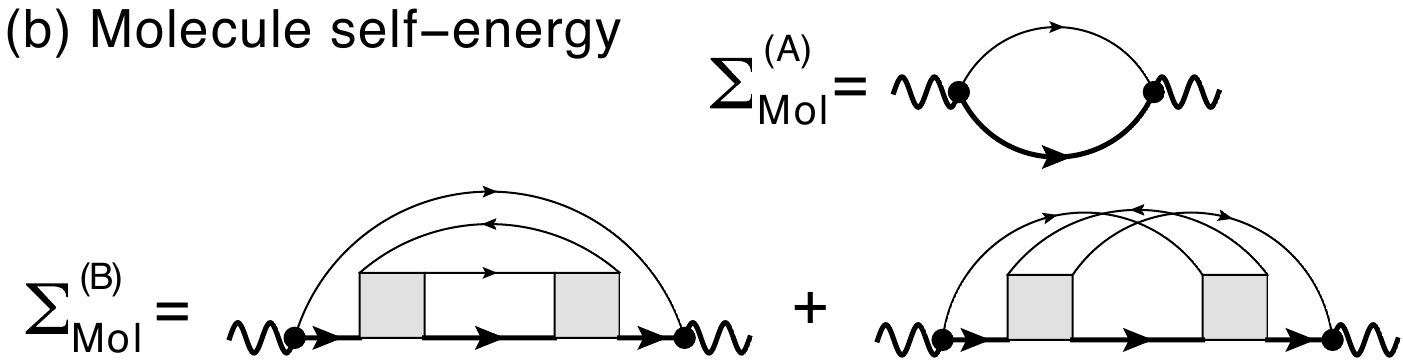}
\caption{{\footnotesize Diagrams showing the 
decay processes A an B for (a) polarons and (b) molecules. The thin lines 
represent majority atoms, wavy lines a molecule, and thick lines a polaron. 
The polaron-molecule matrix element $g$ is represented by a thick dot and the square represents 
off-resonant scattering between a polaron and the majority atoms.}}
\label{fig:feynmandiagrams}
\end{center}
\end{figure} 
 
The phase diagram is animated by considering the rates of each decay 
process. The decay rates of zero momentum polarons and molecules via 
process B are presented in \cite{PhysRevLett.105.020403}. The decay rate of 
a polaron with momentum $p$ through process A is given by the imaginary 
part of the on-shell polaron self-energy $\Sigma_{Pol}^{A}$ shown in 
figure \ref{fig:feynmandiagrams}a, i.e. 
$\Gamma_{Pol}^{A}\left(p\right)=-{\rm Im}\Sigma_{Pol}^{A}\left(p,E_{Pol}+\frac{p^{2}}{2m_{Pol}^{*}}\right)$. 
At zero temperature, one obtains
 \begin{equation}\Gamma_{Pol}^{\left(A\right)}\left(p\right)=-\pi 
Z_{Mol}g^{2}\int_{q<k_{F\uparrow}}\delta\left(\triangle 
E+\frac{p^{2}}{2m_{Pol}^{*}}+\xi_{q}-\frac{\left(p+q\right)^{2}}{2m_{Mol}^{*}}\right),\end{equation}
which is a simple Golden rule expression. Here, 
$g=-\sqrt{\frac{2\pi}{m_{r}^{2}a}}$ is the atom-molecule coupling in 
vacuum  \cite{PhysRevLett.105.020403}, and $m_{r}$ is the atom-molecule reduced mass. To derive 
this equation, we have used a pole expansion of the molecule propagator 
with quasiparticle residue $Z_{Mol}$. The decay rate may be calculated 
analytically. Assuming for simplicity $m_{Mol}^{*}=2m_{\uparrow}$ and 
$m_{Pol}^{*}=m_{\uparrow}$, we obtain
 \begin{equation}\Gamma_{Pol}^{A}\left(p\right)=Z_{Mol}\epsilon_{F\uparrow}\frac{4}{k_{F\uparrow}a}\frac{\left(p-P_{Th}^{A}\right)\left(P_{Th}^{A}+2k_{F\uparrow}-p\right)}{pk_{F\uparrow}},
 \end{equation}
 for $P_{Th}^{A}<p<P_{Th}^{A}+2k_{F\uparrow}$ and zero otherwise. The 
momentum threshold for this process is 
$P_{Th}^{A}=k_{F\uparrow}\left[\sqrt{2\left(1-\frac{\triangle 
E}{\epsilon_{F\uparrow}}\right)}-1\right]$. Note that $P_{Th}^{A}>0$ even 
when $\triangle E>0$, and that one has $\frac{\triangle 
E}{\epsilon_{F\uparrow}}<0.5$ for all scattering lengths 
\cite{PhysRevA.80.053605,PhysRevB.77.020408}.
 
The decay rate of a molecule with momentum $p$ via process A, the creation 
of a polaron and a majority particle, is given by the imaginary part of the 
molecule self-energy $\Sigma_{Mol}^{A}$ depicted in 
figure \ref{fig:feynmandiagrams}b. A calculation analogous to the polaron 
case considered above yields,
 \begin{equation}\Gamma_{Mol}^{A}\left(p\right)=Z_{Pol}\epsilon_{F\uparrow}\frac{2}{k_{F\uparrow}a}\frac{\left(p-P_{Th}^{\tilde{A}}\right)\left(p-P_{Th}^{\tilde{A}}+4k_{F\uparrow}\right)}{pk_{F\uparrow}},
 \end{equation}
with $Z_{Pol}$ the quasiparticle residue for the polaron. The threshold 
momentum for this process is
$P_{Th}^{\tilde{A}}=k_{F\uparrow}\left[2-\sqrt{2\left(1-\frac{\triangle 
E}{\epsilon_{F\uparrow}}\right)}\right]$ with $\Gamma_{Mol}^{A}=0$ for 
$p<P_{Th}^{\tilde{A}}$. Note that in this case, as opposed to the polaron 
case, there is no maximum momentum for the molecule above which there is no 
decay via process A. This is because for large momentum, the molecule can 
always dispose its energy and momentum to a majority particle, whereas the 
polaron has to dispose it into a majority hole within the Fermi sea.
 
The corresponding self-energy for process B, $\Sigma_{Pol}^{B}$ is
shown in figure \ref{fig:feynmandiagrams}a. Generalising the calculations
in \cite{PhysRevLett.105.020403} to non-zero momentum $p$ gives
 \begin{equation}\Gamma_{Pol}^{B}\left(p\right)\propto 
Z_{Mol}\epsilon_{F\uparrow}\left[\frac{\left(\Delta 
E+\frac{p^{2}}{2m_{Pol}^{*}}\right)}{\epsilon_{F\uparrow}}\right]^{\frac{9}{2}}.
\end{equation}
 The analysis in \cite{PhysRevLett.105.020403} showed that the 
proportionality constant in this expression is of order unity.
 
Likewise, the decay rate of a molecule via process B can be obtained
by generalising the calculations in \cite{PhysRevLett.105.020403}
to a finite momentum $p$. We obtain
 \begin{equation}\Gamma_{Mol}^{B}\left(p\right)\propto 
Z_{Pol}\epsilon_{F\uparrow}\left[\frac{\left(-\Delta 
E+\frac{p^{2}}{2m_{Mol}^{*}}\right)}{\epsilon_{F\uparrow}}\right]^{\frac{9}{2}},\end{equation} 
where the constant of proportionality is, again, of order unity. $P_{Th}^{B}$ 
($P_{Th}^{\tilde{B}}$) is the point at which $\Gamma_{Pol}^{B}$ ($\Gamma_{Mol}^{B}$)
equals zero.
 
For a finite momentum, the polaron can scatter off the majority particles 
giving rise to momentum relaxation with a rate $\frac{1}{\tau_{Pol}}$ 
(process C). The high velocity regime $k_{F\downarrow}\ll 
m_{\downarrow}^{*}v\ll k_{F\uparrow}$ analysed in 
\cite{PhysRevLett.100.240406} determines the rate of relaxation of a single impurity.
 If we consider only a single impurity, spin statistics becomes 
redundant, allowing the same calculation to be used for the molecule. The 
momentum relaxation rate, at $T=0$ for molecule or polaron is then,
 \begin{equation}
 \frac{1}{\tau_{Mol/Pol}}=\frac{2\pi}{35}\left|\gamma\right|^{2}\frac{m_{Mol/Pol}^{*3}v^{4}}{k_{F\uparrow}^{2}},
 \end{equation}
 where $\gamma$ is determined by the scattering amplitude 
$U=\frac{\partial\mu_{Mol/Pol}}{\partial 
n_{\uparrow}}=\frac{2\pi^{2}}{m_{\uparrow}k_{F\uparrow}}\gamma$. At 
unitarity, we have $\mu_{Pol}=-\alpha\epsilon_{F\uparrow}$ with 
$\alpha\simeq0.6$ \cite{PhysRevB.77.020408}. One can also use 
$\mu_{Mol}=-\frac{1}{ma^{2}}+\frac{3\pi\tilde{a}}{m}n_{\uparrow}$, valid 
for $\frac{1}{k_{F\uparrow}a}\gtrsim0.7$ \cite{PhysRevB.77.020408} (where 
$\tilde{a}\simeq1.18a$) to find the molecule momentum relaxation rate in 
terms of the interaction strength, 
\begin{equation}\frac{1}{\tau_{Mol}}=\frac{9}{70\pi}\left(k_{F\uparrow}\tilde{a}\right)^{2}\frac{m_{Mol}^{*3}v^{4}}{k_{F\uparrow}^{2}}.
\end{equation} 
The high power of the velocity $v$ and effective mass $m^{*}$ indicate 
impurities at large momenta $p\sim p_{F\uparrow}$ are no longer 
well-defined quasiparticles.
%This becomes increasingly true as the effective mass of a quasiparticle diverges.
 
\begin{figure}
\begin{center}
\includegraphics[width=\columnwidth]{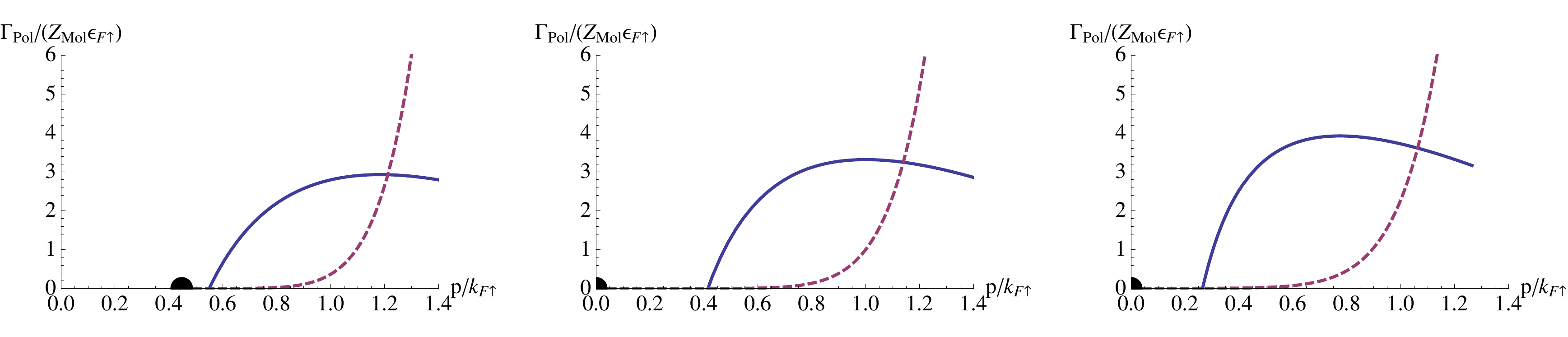}
\includegraphics[width=\columnwidth]{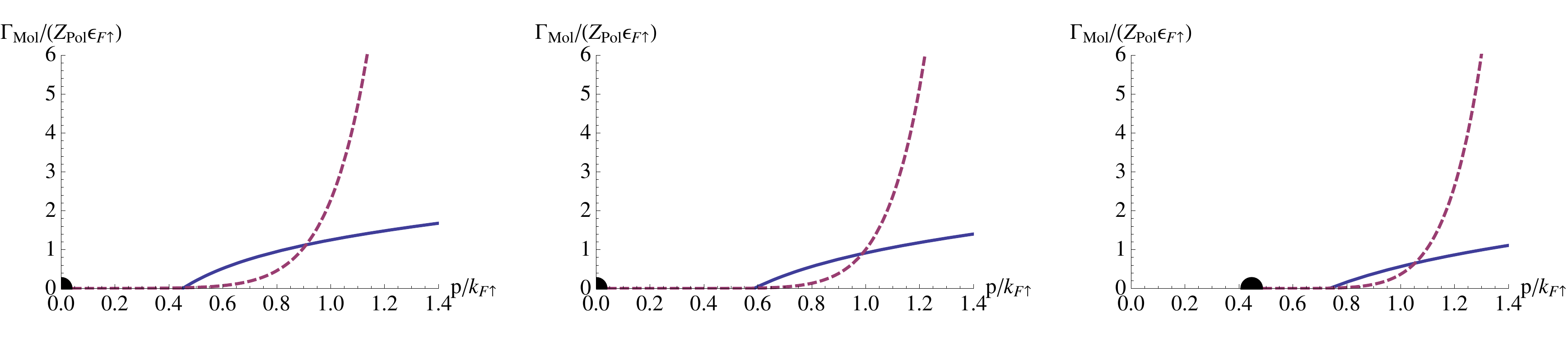}
\caption{{\footnotesize Decay rates of finite 
momentum polarons (top) and molecules (bottom) for processes A (continuous 
line) and B (dashed line). From left to right, we have ${\Delta 
E}/{\epsilon_{F\uparrow}}=-0.2,\:0,\,0.2$. The thick dot 
indicates the threshold momentum for process B, $P_{Th}^{B}$. Process A, above its threshold
dominates over B for $p<p_{F\uparrow}$.}}
\label{fig:Compare Rates}
\end{center}
\end{figure}

In figure \ref{fig:Compare Rates}, we plot the decay rate of polarons and 
molecules via processes A and B as a function of momentum $p$ for various 
$\triangle E$. We see that once $p>P_{Th}^{A}$ and process A sets in, it 
quickly dominates over B. In fact, once active, process A dominates over B 
by several orders of magnitude since the ratio is, to lowest order 
$\frac{\Gamma_{A}}{\Gamma_{B}}\varpropto\left(\frac{p-p_{Th}^{A}}{p_{F\uparrow}}\right)\left(\frac{p_{F\uparrow}}{p_{Th}^{A}}\right)^{10}$. This is expected since process A is a 2-body process, whereas process B is 
a 3-body process. On the other hand, for $p<P_{Th}^{A}$ process B of course 
dominates as process A is not allowed. However, since the typical time 
scale for process B is very long, of order 10-100ms, the momentum 
relaxation of the polaron via process C is in general much faster. This 
justifies our proposed experiment to determine $P_{Th}^{B}$ using a Fermi 
sea of polarons in equilibirum, to suppress process C.

\subsection{Decay rates experiment}

The polaron Fermi sea can also be used to measure the rate of process B. The
initial state is a polaron Fermi sea prepared in the groundstate (i.e. below
the dashed green line in figure \ref{fig:Phase-diagram}). We then instantaneously increase the
interaction strength so that the some fraction of the polarons have a momentum
$p>P_{Th}^{B}$ and so are then susceptible to decay via process B. Alternatively,
by increasing the interaction strength to above ${1}/{(k_{F\uparrow}a)_{c}}$, every polaron
has a momentum $p>P_{Th}^{B}$, and so the whole Fermi sea is susceptible to decay
via process B. The molecules resulting from this decay are unable to decay
via process $\tilde{A}$ or $\tilde{B}$, however they lose momentum via process $\tilde{C}$.
The resultant
state is therefore expected to be a condensate of molecules. A measurement of
the initial growth rate of the number of molecules or the loss rate of polarons
determines the rate of process B for polarons, averaged over the momenta of the polarons decaying, at
a given ${1}/{k_{F\uparrow}a}$.

Similarly, the rate of process A can be measured, with some fraction of the polarons having a momentum $p>P_{Th}^{A}$. In this case, the Fermi sea of polarons will decay via both processes A and B. The two processes can be distinguished since process A has a significantly faster rate and typically results in molecules at finite momentum.

\section{Conclusions}
 
In this paper, we have shown how quasiparticles in a two-component 
Fermi gas behave at finite momenta in the limit of extreme imbalance. By considering energy and momentum
conservation of single impurities, we determined the 
momentum thresholds beyond which the quasiparticles are susceptible
to the most significant decay processes, and we calculated the rates of
each. Using this and assuming 
we can use single quasiparticle decay processes to describe the decay
of a Fermi sea of polaron, we have identified
a region of metastability for the partially polarised normal phase
about the normal to superfluid first order transition. We then described
how the Fermi sea can be used as an experimental probe to observe
the metastable region, determine the crossing point of single impurity
energies $1/(k_{F\uparrow}a)_{c}$ for the first time and measure the decay rate of process A and B. The phase diagram we construct and the adiabatic
experiment we propose to explore it can be used as a measure of the
unknown molecule-polaron interaction strength, and leaves open the possibility
of observing a novel state; a mixture of molecules and polarons.

\ack{K.S. and C.L. acknowledge support from the EPSRC through the Advanced Fellowship EP/E053033/1.
P.M. acknowledges support from the ESF/MEC project FERMIX
(FIS2007-29996-E), the Spanish MEC project FIS2008-01236, and the
Catalan project 2009-SGR-985. A.R. also acknowledges support from the ESF/MEC project FERMIX
(FIS2007-29996-E).}

\section*{References}
\bibliographystyle{unsrt}
\bibliography{sade1201.bib}

\end{document}